\documentclass[journal=jacsat,manuscript=article]{achemso}

\usepackage{subfig, graphicx}
\usepackage{verbatim}
\usepackage[version=3]{mhchem}
\usepackage{latexsym}
\usepackage{amsmath}
\usepackage{setspace}
\usepackage[nolist]{acronym}
\usepackage{multirow}
\usepackage{upgreek}
\usepackage{amssymb}
\usepackage{tipa}
\usepackage{dcolumn}
\usepackage{bm}
\usepackage{newunicodechar} 
\usepackage{tabularx}
\usepackage{makecell}
\usepackage{siunitx}
\usepackage{xcolor}
\usepackage{soul}

\author{Xinwei Wang}
\affiliation{ 
Department of Materials, Imperial College London, Exhibition Road, London SW7 2AZ, UK
}
\author{Seán R. Kavanagh}
\affiliation{ 
Center for the Environment, Harvard University, 29 Oxford St, Cambridge, MA 02138, US
}

\author{Aron Walsh}
 \email{a.walsh@imperial.ac.uk}
\affiliation{ 
Department of Materials, Imperial College London, Exhibition Road, London SW7 2AZ, UK
}
\alsoaffiliation{
Department of Physics, Ewha Womans University, 52 Ewhayeodae-gil, Seodaemun-gu, Seoul, 03760, South Korea
}

\title
  {Sulfur Vacancies Limit the Open-circuit Voltage of \ce{Sb2S3} Solar Cells}

\begin{document}

\begin{acronym}
\acro{PV}{photovoltaic}
\acro{PVs}{photovoltaics}
\acro{PCE}{power conversion efficiency}
\acro{PCEs}{power conversion efficiencies}
\acro{1D}{one-dimensional}
\acro{VASP}{Vienna Ab initio Simulation Package}
\acro{DFT}{density functional theory}
\acro{PAW}{projector augmented-wave}
\acro{TLs}{transition levels}
\acro{SRH}{Shockley-Read-Hall}
\acro{PESs}{potential energy surfaces}
\acro{CBM}{conduction band minimum}
\acro{VBM}{valence band maximum}
\acro{cc}{configuration coordinate}
\acro{CB}{conduction band}
\acro{VB}{valence band}
\end{acronym}

\begin{abstract}
Antimony sulfide (\ce{Sb2S3}) is a promising candidate as an absorber layer for single-junction solar cells and the top subcell in tandem solar cells. However, the power conversion efficiency of \ce{Sb2S3}-based solar cells has remained stagnant over the past decade, largely due to trap-assisted non-radiative recombination. Here we assess the trap-limited conversion efficiency of \ce{Sb2S3} by investigating non-radiative carrier capture rates for intrinsic point defects using first-principles calculations and Sah-Shockley statistics. Our results show that sulfur vacancies act as effective recombination centers, limiting the maximum efficiency of \ce{Sb2S3} to 16\% light to electricity. The equilibrium concentrations of sulfur vacancies remain relatively high regardless of growth conditions, indicating the intrinsic limitations imposed by these vacancies on the performance of \ce{Sb2S3}.
\end{abstract}

Antimony sulfide (\ce{Sb2S3}) has attracted great research interest as an emerging light-absorbing material for next-generation \ac{PV} devices, driven by its earth-abundant and environmental-friendly constituents, as well as its attractive optical and electronic properties \cite{kondrotas2018sb2s3}.
\ce{Sb2S3} has a high optical absorption coefficient (\textgreater{ }\SI{1e4}{\per\cm} in the visible region \cite{ghosh1979optical}), decent carrier mobility \cite{chalapathi2020influence,wang2022band}, and excellent thermal and chemical stability \cite{wang2023perspective}.
Its band gap of \SIrange[range-units=single, range-phrase=--]{1.7}{1.8}{\electronvolt} \cite{kondrotas2018sb2s3} is well-aligned with the spectra of indoor light sources and is ideal for the top subcell in tandem solar cells.
Additionally, its relatively low melting point (\SI{550}{\degreeCelsius} \cite{kondrotas2018sb2s3}) facilitates the growth of high-quality \ce{Sb2S3} crystalline films at moderate temperatures.
Despite these advantages, the \ac{PCEs} of single-junction \ce{Sb2S3} solar cells remain low.
The highest recorded efficiencies are 8.3\% for the planar geometry type \cite{zhu2023parallel} and 7.5\% for sensitized type \cite{choi2014highly}, far below the thermodynamic limit of $\sim$30\% for a material with this band gap  \cite{shockley1961detailed}.

The main challenge impeding further efficiency improvements in \ce{Sb2S3} solar cells is the significant open-circuit voltage (\textit{V}$_\textrm{OC}$) deficit.
Despite various device architectures and fabrication strategies \cite{wang2023perspective}, the \textit{V}$_\textrm{OC}$ deficit for the most efficient \ce{Sb2S3} devices remain greater than \SI{0.9}{\volt}\cite{zhu2023parallel}, indicating a high electron-hole recombination rate.
The detailed balance principle\cite{shockley1961detailed} predicts a minimum \textit{V}$_\mathrm{OC}$ deficit (defined as $E_\mathrm{g}/\textit{q}-V\mathrm{^{SQ}_{OC}}$) of $\sim$\SI{0.27}{\volt} due to unavoidable band-to-band radiative recombination at \SI{300}{\kelvin} for a material with a band gap of \SI{1.7}{\eV}\cite{nayak2019photovoltaic}.
Besides band-to-band recombination, electron and hole capture processes can occur non-radiatively through multiple-phonon emission (either via defect-mediated processes or the Auger–Meitner effect \cite{stoneham1975theory,matsakis2019renaming}), or radiatively via defect-mediated pathways that involve the emission of photons.
For defect-mediated capture processes, radiative capture cross-sections are typically on the order of $10^{-5} - 10^{-4} \, \si{\angstrom\squared}$, which are significantly smaller than the $10^{-2} - 10^{4} \, \si{\angstrom\squared}$ range for non-radiative processes.
The Auger–Meitner process \cite{matsakis2019renaming} becomes relevant only in systems with exceptionally high defect and carrier concentrations (usually \textgreater \SI{e17}{\per\cubic\cm}) \cite{stoneham1975theory}.
Consequently, defect-mediated non-radiative recombination (\ac{SRH} recombination) is widely recognized as the dominant loss mechanism in \ce{Sb2S3} solar cells \cite{chen2020open}.

There is ongoing debate regarding whether \ac{SRH} recombination predominantly occurs at the surface/interface or within the bulk of \ce{Sb2S3} \cite{boix2012hole}, and whether vacancies or antisites are the most detrimental defects \cite{wang2022novel}.
There have also been predictions around multi-carrier trapping.\cite{liu2023strong}
Identifying the specific type of defect can be challenging when relying solely on experimental methods, making complementary theoretical simulations essential.
Previous studies on defect identification in \ce{Sb2S3} have primarily focused on comparing energy levels and defect types (acceptor vs. donor).
However, due to the low crystal symmetry of \ce{Sb2S3}, there are multiple types of defects with numerous trap states distributed across the band gap.
Furthermore, not all deep-level traps are active recombination centers capable of rapid carrier capture.
A more comprehensive understanding of charge carrier recombination kinetics through defects is therefore essential.
Traditional \ac{SRH} theory \cite{shockley1952statistics,hall1952electron} is commonly used to calculate recombination rates via single-level defects.
Nevertheless, many studies in this field continue to rely on \ac{SRH} theory with additional approximations for defects with multiple energy levels. 
Some researchers neglect the carrier re-emission processes and derive effective total carrier capture coefficients \cite{alkauskas2016role}, while others treat each defect level independently and sum the \ac{SRH} recombination rates for individual single-level defects \cite{kim2020upper,kim2021ab}.
These approximations, however, can lead to significant errors in systems with correlated defect transitions or negative correlation energies \cite{willemen1998modelling,steingrube2012limits}, such as antimony chalcogenides \cite{wang2023four}.
For such systems, Sah-Shockley statistics \cite{sah1958electron}, which account for multiple defect levels, provide more accurate predictions.

In this work, we have performed systematic first-principles calculations to investigate intrinsic point defects in \ce{Sb2S3} using a global structure searching approach \textsc{ShakeNBreak} \cite{mosquera2022shakenbreak,mosquera2023identifying,wang2023four}.
We further assessed non-radiative carrier recombination via these defects using Sah-Shockley statistics \cite{sah1958electron}.
By accounting for both band-to-band radiative and trap-mediated non-radiative recombination, we predict the upper limit of \ac{PCE} in \ce{Sb2S3}.
Our results reveal that sulfur vacancies are the most detrimental defects, contributing significantly to \textit{V}$_\textrm{OC}$ loss due to their consistently high equilibrium concentrations under a range of growth conditions.

\begin{figure}[t!]
    \centering
    {\includegraphics[width=0.8\textwidth]{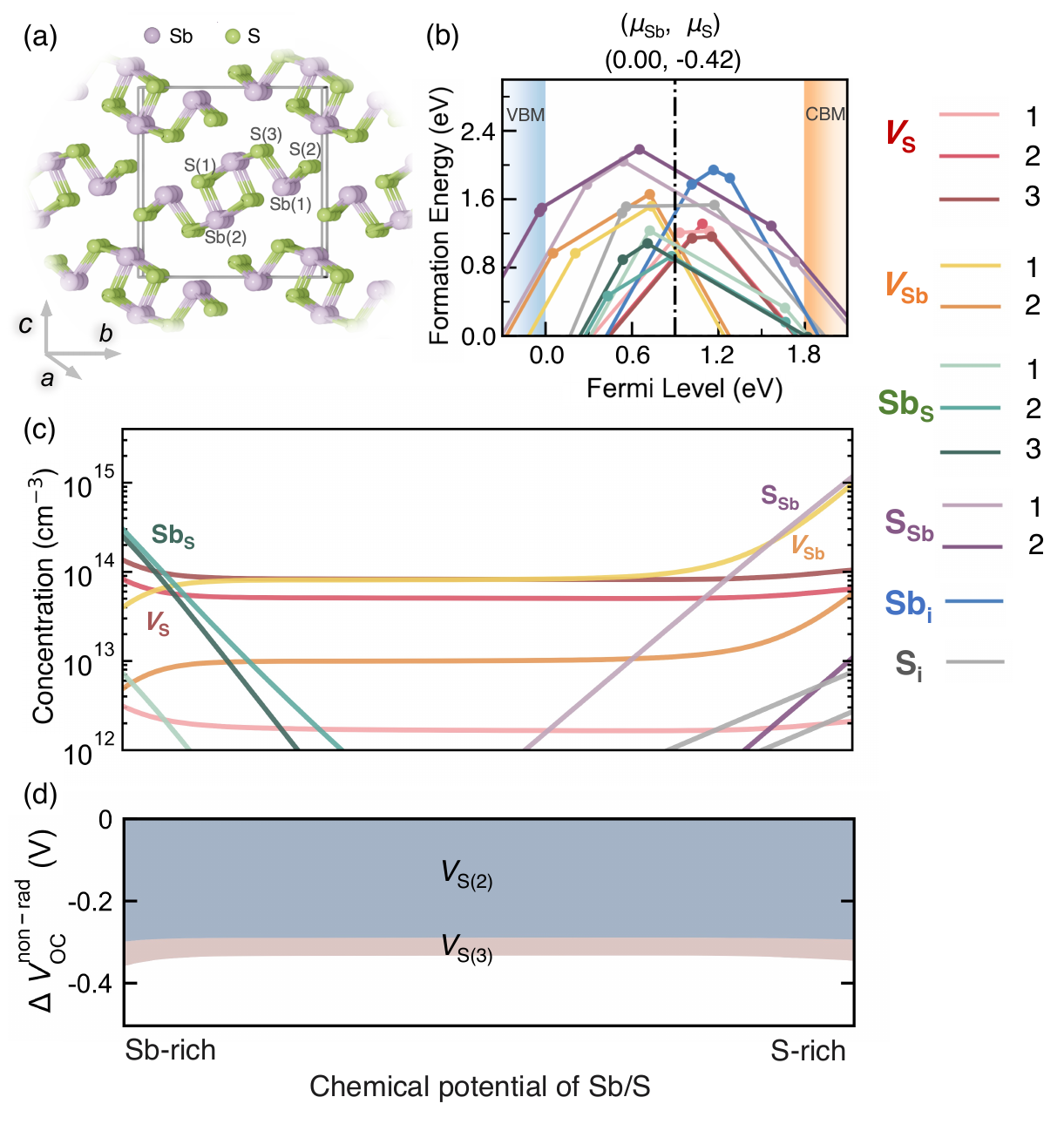}} \\
     \caption{(a) Ground-state crystal structure (\textit{Pnma} space group) of \ce{Sb2S3}. The crystallographic unit cell is represented by a cuboid. Inequivalent sites are denoted by the atom labels enclosed in parentheses. (b) Calculated formation energies of intrinsic point defects in \ce{Sb2S3} under Sb-rich growth conditions 
     using \textsc{doped}\cite{kavanagh2024} and \textsc{ShakeNBreak}\cite{mosquera2022shakenbreak}.
     The slopes of solid lines represent charge states, while the filled circles indicate thermodynamic transition levels. The valence band maximum (VBM) is set to \SI{0}{eV}, and the conduction band minimum (CBM) is obtained from the calculated fundamental (indirect) band gap of \SI{1.79}{eV} by the HSE06 functional. 
     Vertical dashed lines indicate self-consistent Fermi levels. The legend numbers correspond to different inequivalent sites; for interstitials, only the lowest energy states at each charge state are shown. (c) Equilibrium defect concentration at \SI{300}{\kelvin} in \ce{Sb2S3} crystals grown at \SI{603}{\kelvin} \cite{zhu2023parallel,chen2019preferentially} as a function of growth condition. (d) \textit{V}$\mathrm{_{OC}}$ deficit contributed by non-radiative recombination ($\Delta$\textit{V}$\mathrm{_{OC}^\mathrm{non-rad}}$) in \ce{Sb2S3} as a function of growth condition, decomposed into individual defect contributions. $\Delta$\textit{V}$\mathrm{_{OC}^\mathrm{non-rad}}$ is defined as the difference between the values of \textit{V}$\mathrm{_{OC}}$ and \textit{V}$\mathrm{_{OC}^\mathrm{rad}}$. Defect species with $\Delta$\textit{V}$\mathrm{_{OC}^\mathrm{non-rad}}$ $\textless$ 0.05 V are not shown. Film thickness is assumed to be 400 nm \cite{zhu2023parallel,chen2019preferentially}.}
    \label{fig_e}
\end{figure}

\ce{Sb2S3} forms an orthorhombic crystal structure and belongs to the \textit{Pnma} space group \cite{hofmann1933struktur} (as shown in Fig. \ref{fig_e}(a)).
The structure consists of quasi-\ac{1D} [Sb$_4$S$_6$]$_n$ ribbons along the [100] direction, which are linked together by weak interactions\cite{wang2022lone}. 
The low crystal symmetry of the structure results in distinct coordination environments for each Sb and S element within the unit cell, leading to two inequivalent Sb sites and three inequivalent S sites, all of which were considered in our calculations. 

\textbf{Equilibrium bulk defects.} All intrinsic point defects, including vacancies, antisites, and interstitials, were systematically investigated by first-principles calculations.
Details of defect generation and optimization are provided in Methods.
The defect formation energy diagram in Fig. \ref{fig_e}(b) plots the thermodynamically stable charge states as a function of the Fermi level (E$_F$) within the band gap under Sb-rich conditions.
Similar to previous studies on \ce{Sb2Se3} \cite{wang2024upper}, all intrinsic point defects in \ce{Sb2S3} exhibit amphoteric behavior, with stable positive and negative charge states depending on the position of E$_F$.
This suggests strong charge compensation, which reduces carrier density and leads to poor electrical
conductivity, consistent with other reports \cite{cardenas2009carbon}.
All point defects with low formation energies have deep thermodynamic \ac{TLs}, making it challenging to identify detrimental defects based solely on their deep-level characteristics.

The equilibrium defect concentration as a function of chemical potential is further calculated under the constraint of charge neutrality \cite{buckeridge2019equilibrium}.
As illustrated in Fig. \ref{fig_e}(c), under Sb-rich conditions, Sb$_\mathrm{S}$ and \textit{V}$_\mathrm{S}$ are dominant defects with high concentrations (\textgreater \SI{e14}{\per\cubic\cm}).
As the sulfur chemical potential ($\mu_\textrm{S}$) increases, the density of Sb$_\mathrm{S}$ decreases significantly, while that of S$_\mathrm{Sb}$ rises sharply.
In contrast, the variation in vacancy concentration with $\mu_\textrm{S}$ is less pronounced.
\textit{V}$_\mathrm{S}$ maintains a consistently high concentration across various growth conditions. 
While the concentration of \textit{V}$_\mathrm{Sb}$ initially increases slowly with $\mu_\textrm{S}$, it then increases dramatically as the system approaches S-rich conditions, ultimately reaching a high concentration under S-rich conditions.
The insensitivity of defect concentrations to growth conditions can be explained by defect-correlations,\cite{huang2021more} with Schottky-type disorder between \textit{V}$_\mathrm{S}$ and \textit{V}$_\mathrm{Sb}$ leading to charge compensation across most of the chemical potential range. 
All interstitials are found to have low concentrations, which agrees with the experimental observation that interstitials have a negligible impact on carrier lifetime \cite{lian2021revealing}.


We note that previous first-principles studies on \ce{Sb2S3} \cite{cai2020intrinsic,zhao2021intrinsic,huang2021more} commonly reported sulfur vacancies as donors and antimony vacancies as acceptors, rather than amphoteric defects.
This discrepancy likely stems from the absence of global structure-searching methods and a limited exploration of charge states in these earlier studies. 
Therefore, our findings emphasize the importance of using global structure searching for accurate defect predictions in chalcogenide semiconductors.

\textbf{Carrier capture under steady-state illumination.} 
The defect-mediated carrier capture processes via multi-phonon emission and corresponding recombination kinetics in \ce{Sb2S3} were then investigated.
The electron and hole capture coefficients for defects with high carrier concentrations were calculated following the procedure outlined in our previous work \cite{wang2024upper}.
The complete pathways for electron and hole capture can be found in Fig. S3.
The maximum achievable conversion efficiency is further predicted to quantify the impact of trap-assisted electron-hole recombination on the performance of \ce{Sb2S3} solar cells.
The predicted open-circuit voltage \textit{V}$\mathrm{_{OC}}$ deficit due to radiative recombination is \SI{0.10}{/V}.
Non-radiative recombination contributes significantly to the total \textit{V}$\mathrm{_{OC}}$ deficit, with values of \SI{0.45}{/V} and \SI{0.44}{/V} under Sb-rich and S-rich conditions, respectively (Fig. S4(a)).
Further analysis of the \textit{V}$\mathrm{_{OC}}$ loss due to non-radiative recombination ($\Delta$\textit{V}$\mathrm{_{OC}^\mathrm{non-rad}}$) shows that the highest loss of \SI{0.36}{V} occurs under Sb-rich conditions, with losses ranging from 0.33 to \SI{0.36}{V} across different growth conditions (Fig.  \ref{fig_e}(d)).
To identify the most detrimental defect species, the contributions to $\Delta$\textit{V}$\mathrm{_{OC}^\mathrm{non-rad}}$ are divided by individual defect type.
As shown in Fig. \ref{fig_e}(d), the conversion efficiency of \ce{Sb2S3} is primarily limited by sulfur vacancies, whereas antimony vacancies, antisites and interstitials have negligible impact on non-radiative recombination. 
Among the sulfur vacancies, \textit{V}$_\mathrm{S(2)}$ and \textit{V}$_\mathrm{S(3)}$ are found to be the most harmful, owing to their high concentrations and large carrier capture coefficients for both electrons and holes.
Minimizing these defects is thus crucial for improving the \ac{PCE} of \ce{Sb2S3} solar cells.

\begin{figure}[h!]
    \centering    {\includegraphics[width=\textwidth]{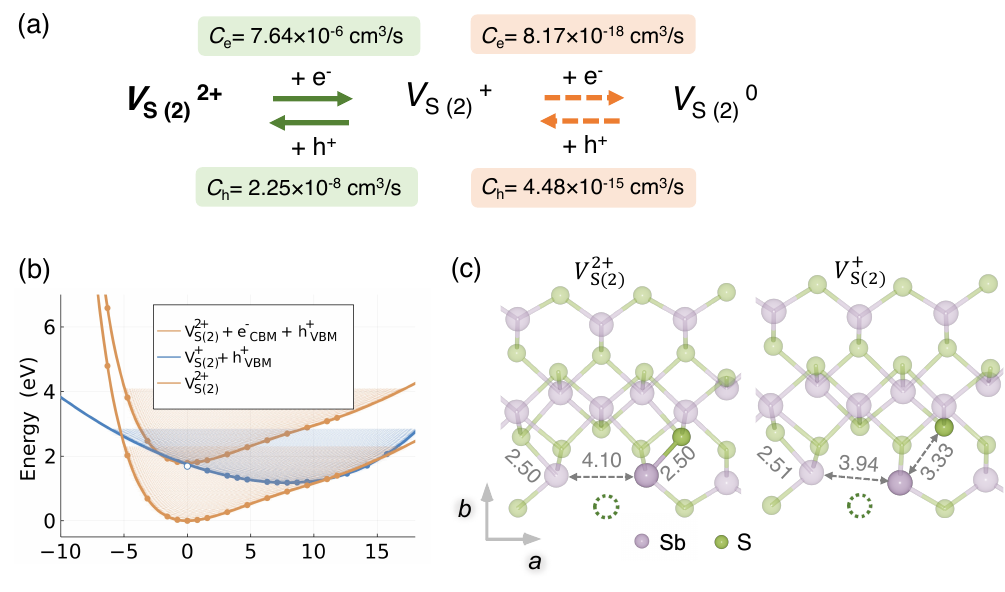}} \\
    \caption{(a) Pathways for electron and hole capture by \textit{V}$\mathrm{_{S(2)}}$. $C_e$ and $C_h$ are electron and hole capture coefficients, respectively, calculated using \textsc{CarrierCapture}\cite{kim2020carriercapture}.  Green and orange indicate rapid and slow capture processes, respectively. (b) One-dimensional (1D) configuration coordinate diagram for charge transitions between \textit{V}$\mathrm{^{2+}_{S(2)}}$ and \textit{V}$\mathrm{^{+}_{S(2)}}$. Solid circles are data points obtained by DFT calculations and used for fitting, while hollow circles are discarded for fitting due to charge delocalisation. Solid lines represent best fits to the data. (c)   Defect configurations of \textit{V}$\mathrm{^{2+}_{S(2)}}$ and \textit{V}$\mathrm{^{+}_{S(2)}}$. The bond lengths in Å are labelled, and the vacant S site is denoted by a dotted circle.}
    \label{fig_pes}
\end{figure}

As previously discussed, the primary defect species contributing to the largest \textit{V}$\mathrm{_{OC}}$ loss is predicted to be \textit{V}$_\mathrm{S(2)}$, an amphoteric defect with multiple accessible charge states: 0, $\pm$1 and $\pm$2.
The overall recombination rate of electrons and holes via such a multi-valent defect is determined by competing transitions between different charge states, rather than by the sum of individual transitions.
Consequently, the commonly used \ac{SRH} statistics, which address only the transition between two charge states may be inadequate so we also implement a treatment based on Sah-Shockley statistics which includes transitions between multiple charge states.\cite{sah1958electron}

For carrier capture transitions involving \textit{V}$_\mathrm{S(2)}$, the \textit{V}$^{2+}_\mathrm{S(2)}$ state is considered as the starting point, since it has the highest equilibrium concentration under various growth conditions (Fig. \ref{fig_e}(b) and Fig. S1).
The detailed carrier capture transition pathways associated with \textit{V}$_\mathrm{S(2)}$ are illustrated in Fig. \ref{fig_pes}(a).
In non-radiative carrier recombination processes, \textit{V}$^{2+}_\mathrm{S(2)}$ captures an electron from the \ac{CBM}, followed by hole capture by \textit{V}$^{+}_\mathrm{S(2)}$ from the \ac{VBM}.
These capture processes can be described using a \ac{cc} diagram. 
As shown in Fig. \ref{fig_pes}(b), the \ac{PESs} of \textit{V}$^{2+}_\mathrm{S(2)}$ and \textit{V}$^{+}_\mathrm{S(2)}$ are plotted as a function of \ac{1D} generalised coordinate \textit{Q}, which represents atomic displacement (Fig. \ref{fig_pes}(b)).
The coordinate \textit{Q} is generated by linearly interpolating between the ground-state configurations of \textit{V}$^{2+}_\mathrm{S(2)}$ and \textit{V}$^{+}_\mathrm{S(2)}$, and it corresponds to the vibrations most strongly coupled to the structural distortion during the transition.
The validity of this \ac{1D} approximation is supported by the linear fit of the wavefunction overlap $\langle{\psi_{i}}|{\psi_{f}}\rangle$ as a function of \textit{Q} (shown in Fig. S5).

Table \ref{table_pes} summarises the carrier capture coefficients and cross-sections at room temperature, along with key parameters used in the calculations performed within \textsc{CarrierCapture}.\cite{kim2020carriercapture} 
The large electron-phonon matrix element $W_{if}$ shows the strong promoting character of the configuration coordinate.
The mass-weighted displacement $\Delta Q$ quantifies the structural difference between the two defect charge states involved in the capture process.
For the transition between \textit{V}$^{2+}_\mathrm{S(2)}$ and \textit{V}$^{+}_\mathrm{S(2)}$, the main contribution to $\Delta Q$ of 7.88 amu$^{1/2}${\AA } arises from the shortening (lengthening) of a Sb-S bond adjacent to \textit{V}$_\mathrm{S(2)}$ during the electron (hole) capture process (Fig. \ref{fig_pes}(c)).
The \ac{PESs} were generated by interpolating between the equilibrium structures of \textit{V}$^{2+}_\mathrm{S(2)}$ and \textit{V}$^{+}_\mathrm{S(2)}$ using single-point \ac{DFT} calculations (Fig. \ref{fig_pes}(b)). 
During the non-radiative electron capture process by \textit{V}$^{2+}_\mathrm{S(2)}$, the initial (excited) state is represented by the uppermost orange curve, while the final (ground) state corresponds to the blue curve. 
The two PESs intersect at $\Delta$\textit{E}$_\textrm{b}$ = \SI{5}{\meV} above the minimum of the excited state.
This small $\Delta$\textit{E}$_\textrm{b}$, combined with a large phonon overlap, results in a large electron capture coefficient ($\textit{C}_e$) of \SI{7.64e-6}{\cubic\cm\per\second} at room temperature.
For hole capture by \textit{V}$^{+}_\mathrm{S(2)}$, the initial and final states are represented by the blue and bottom-most orange curves, respectively. 
The weaker Coulomb repulsion between holes and \textit{V}$^{+}_\mathrm{S(2)}$, the reduced pathway degeneracy $g$ and a larger $\Delta$\textit{E}$_\textrm{b}$ of \SI{121}{\meV} (Table \ref{table_pes}), contribute to a smaller hole capture coefficient ($C_h$) of \SI{2.25e-8}{\cubic\cm\per\second} at room temperature.
Subsequent electron capture by \textit{V}$\mathrm{^{0}_{S(2)}}$ or hole capture by \textit{V}$\mathrm{^{+}_{S(2)}}$ proceeds much more slowly, with capture coefficients $>$ \SI{1e-15}{\cubic\cm\per\second} (Fig. \ref{fig_pes}(c)).
Therefore, the \textit{V}$\mathrm{^{2+}_{S(2)}}$ $\rightleftarrows$ \textit{V}$^{+}_\mathrm{S(2)}$ recombination cycle is efficient, making the overall electron-hole recombination process at \textit{V}$_\mathrm{S(2)}$ primarily limited by the hole capture process \textit{V}$\mathrm{^{+}_{S(2)}} + \textrm{h}^+ \rightarrow$ \textit{V}$\mathrm{^{2+}_{S(2)}}$. 

\begin{table*}[ht!]
\caption{Key parameters used to calculate the carrier capture coefficients in the transition of \textit{V}$^{2+}_\mathrm{S(2)}$ $\leftrightarrow$ \textit{V}$^{+}_\mathrm{S(2)}$: mass-weighted distortion $\Delta$\textit{Q} (amu$^{1/2}$\AA), energy barrier $\Delta$\textit{E}$_\textrm{b}$  (meV), degeneracy factor \textit{g} of the final state, electron-phonon coupling matrix element $W_{if}$ and scaling factor \textit{s}(\textit{T})\textit{f} at \SI{300}{\kelvin}, along with calculated capture coefficient \textit{C} (\SI{}{\cubic\cm\per\second}) and cross-section $\sigma$ (\SI{}{\cm\squared}) at \SI{300}{\kelvin}}
\label{table_pes}
\vspace{10pt}
 \small
\begin{tabular*}{\textwidth}{@{\extracolsep{\fill}}c@{\extracolsep{\fill}}c@{\extracolsep{\fill}}c@{\extracolsep{\fill}}c@{\extracolsep{\fill}}c@{\extracolsep{\fill}}c@{\extracolsep{\fill}}c@{\extracolsep{\fill}}c@{\extracolsep{\fill}}c}
    \hline
Species & $\Delta$\textit{Q} & \begin{tabular}[c]{@{}c@{}}Capture\\ process\end{tabular} & $\Delta$\textit{E}$_\textrm{b}$ & \textit{g} & $W_{if}$ & \textit{s}(\textit{T})\textit{f} & \textit{C} & $\sigma$ \\     \hline
\multirow{2}{*}{\textit{V}$_\mathrm{S}$} &\multirow{2}{*}{7.88} & Electron & \SI{5}{} & 4 & \SI{1.65e-2}{} &  5.34 &\SI{7.64e-6}{} & \SI{4.14e-13}{} \\
&  & Hole & \SI{121}{} & 1 & \SI{3.22e-2}{} & 0.36 &\SI{2.25e-8}{} & \SI{1.54e-15}{} \\     \hline
\end{tabular*}
\end{table*}

In conclusion, the low carrier concentrations and low open-circuit voltage (\textit{V}$\mathrm{_{OC}}$) in \ce{Sb2S3}-based solar cells are linked to intrinsic point defects.
The amphoteric nature of these defects leads to strong charge compensation and thus reduced carrier concentrations.
The accessibility of multiple charge states for a single defect species is dealt with using Sah-Shockley statistics.
Vacancies and antisites emerge as the most prevalent defects, with concentrations \textgreater \SI{e12}{\per\cubic\cm}.
Among these, sulfur vacancies are identified as the most detrimental, contributing substantially to the \textit{V}$\mathrm{_{OC}}$ deficit.
Our calculations show that band-to-band radiative and trap-mediated non-radiative recombination result in a \textit{V}$\mathrm{_{OC}}$ loss up to \SI{0.45}{/V}, limiting the \ac{PCE} to 16\% under Sb-rich conditions.
Notably, the equilibrium concentrations of key recombination centers are steadily high across various growth conditions due to defect-correlation effects, highlighting the challenge in mitigating these defects.
Therefore, effective defect engineering is crucial to improve the performance of \ce{Sb2S3} solar cells.
While strategies to eliminate the harmful effects of sulfur vacancies remain an open question, studies suggest that oxygen or selenium may passivate these vacancies \cite{cai2020intrinsic,zhang2022sulfur}, though care must be taken to avoid the formation of secondary phases that could degrade performance \cite{choi2014highly}. 
Additionally, post-sulfurization treatments have been shown to enhance crystallinity and reduce recombination losses \cite{luo2020fabrication,choi2014highly}. Further research into optimized defect passivation techniques is necessary to unlock the full potential of \ce{Sb2S3}-based solar cells.

\subsection{Methods}

\textbf{Electron-hole recombination statistics.}
We predict the \ac{PCE} of a single-junction solar cell by incorporating both radiative (band-to-band) and defect-mediated non-radiative recombination losses.
The radiative limit is calculated using the bandgap, film thickness-dependent optical absorption, the standard AM1.5 solar spectrum, and an operating temperature of \SI{300}{\kelvin}. This follows the methodology developed by Kim et al \cite{kim2020upper,kim2021ab}. The defect-mediated recombination rate is influenced by three primary factors: carrier capture coefficients, defect concentration, and recombination statistics \cite{das2020deep}. Here, we focus on the statistical modeling of recombination processes with other details provided as Supporting Information. 

The foundational theory of recombination via single-level defects was first established by Shockley, Read \cite{shockley1952statistics} and Hall \cite{hall1952electron}. 
Sah and Shockley extended the statistics for defects with multiple energy levels \cite{sah1958electron}.
The key difference is that Sah-Shockley theory accounts for correlated transitions between defects with different charge states, while the \ac{SRH} model treats them as independent. The amphoteric model is reported to be necessary for systems with negative correlation energies \cite{willemen1998modelling} such as antimony chalcogenides \cite{wang2023four}.

\begin{figure}[h!]
    \centering    {\includegraphics[width=0.5\textwidth]{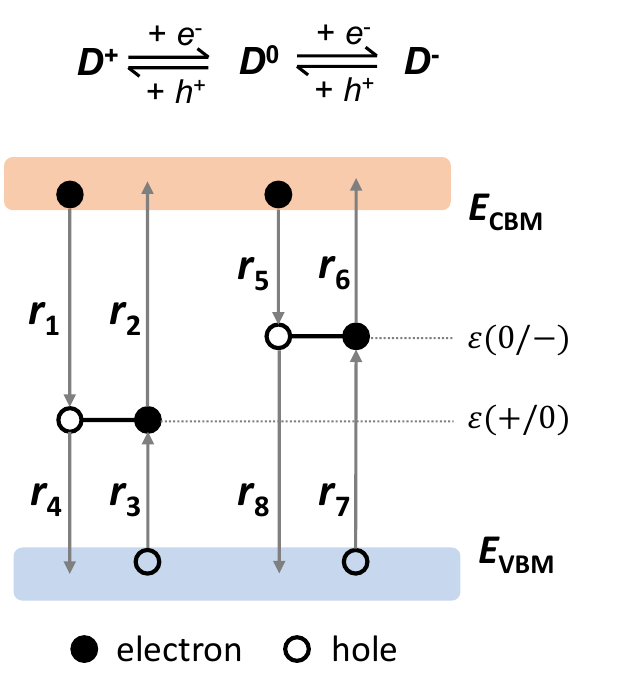}} \\
    \caption{Schematic diagram of capture and emission processes for an amphoteric defect with two transition levels $\varepsilon(+/0)$ and $\varepsilon(0/-)$. Details of rates $r_1$ - $r_8$ are provided in Table \ref{tab_recom}.}
    \label{fig_recom}
\end{figure}

We illustrate the recombination statistics with an amphoteric defect, which can exist in three charge states: $\textit{D}^+$, $\textit{D}^0$, and $\textit{D}^-$.
The net recombination depends on eight individual (i.e. four capture and four emission) processes between these three states as shown in Fig. \ref{fig_recom}.
The corresponding capture and emission rates are given in Table \ref{tab_recom}.

~\\
\begin{table*}[h!]
\centering
 \caption{Capture and emission processes in an amphoteric defect. \textit{n} and \textit{p} are concentrations of electrons and holes, respectively. $\textit{C}_{n/p}$ and $\textit{e}_{n/p}$ are capture and emission coefficients for electrons/holes, respectively. The superscript refers to the starting charge state of the process. \textit{N}$_\textrm{T}$ is the total concentration of defects. \textit{F} is the occupation probability at a certain charge state}
 \label{tab_recom}
 \vspace{10pt}
 \begin{tabular*}{0.75\textwidth}{@{\hspace{0.5cm}\extracolsep{\fill}}c@{\extracolsep{\fill}}c@{\extracolsep{\fill}}c@{\hspace{0.1cm}\extracolsep{\fill}}}
\hline
Process & Transition & Rate \\ \hline
\multicolumn{1}{r}{\hspace{0.8cm}Electron capture $r_1$} & $\textit{D}^+$ + $e^-$ $\rightarrow$ $\textit{D}^0$ & \multicolumn{1}{l}{\textit{n}$\textit{C}^+_n$ $\textit{N}_T$ $\textit{F}^+$\hspace{1cm}}\\
\multicolumn{1}{r}{Electron emission $r_2$} & $\textit{D}^0$ $\rightarrow$ $\textit{D}^+$ + $e^-$ & \multicolumn{1}{l}{$\textit{e}^0_n$$\textit{N}_T$ $\textit{F}^0$}\\
\multicolumn{1}{r}{Hole capture $r_3$} & $\textit{D}^0$ + $h^+$ $\rightarrow$ $\textit{D}^+$ & \multicolumn{1}{l}{\textit{p}$\textit{C}^0_p$ $\textit{N}_T $$\textit{F}^0$}\\
\multicolumn{1}{r}{Hole emission $r_4$} & $\textit{D}^+$ $\rightarrow$  $\textit{D}^0$ + $h^+$ & \multicolumn{1}{l}{$\textit{e}^+_p$$\textit{N}_T$$ \textit{F}^+$}\\
\multicolumn{1}{r}{Electron capture $r_5$} & $\textit{D}^0$ + $e^-$ $\rightarrow$ $\textit{D}^-$ & \multicolumn{1}{l}{\textit{n}$\textit{C}^0_n$ $\textit{N}_T$ $\textit{F}^0$}\\
\multicolumn{1}{r}{Electron emission $r_6$} & $\textit{D}^-$ $\rightarrow$ $\textit{D}^0$ + $e^-$ &\multicolumn{1}{l}{$\textit{e}^-_n$$\textit{N}_T$ $\textit{F}^-$} \\
\multicolumn{1}{r}{Hole capture $r_7$} & $\textit{D}^-$ + $h^+$ $\rightarrow$  $\textit{D}^0$ & \multicolumn{1}{l}{\textit{p}$\textit{C}^-_p$ $\textit{N}_T$ $\textit{F}^-$}\\
\multicolumn{1}{r}{Hole emission $r_8$} & $\textit{D}^0$ $\rightarrow$ $\textit{D}^-$ + $h^+$ & \multicolumn{1}{l}{$\textit{e}^0_p$ $\textit{N}_T$ $\textit{F}^0$}\\\hline
\end{tabular*}
\end{table*}
\clearpage
Emission coefficients are derived from the principle of detailed balance \cite{shockley1952statistics}
\begin{equation}
\begin{aligned}
\textit{e}^0_n&=\frac{1}{\textit{g}}\textit{C}_n^+\textit{N}_\textrm{C}\textrm{exp}[\frac{\varepsilon(+/0)-\textit{E}_\textrm{C}}{k_B\textit{T}}] \\
\textit{e}^-_n&=\textit{g}\textit{C}_n^0\textit{N}_\textrm{C}\textrm{exp}[\frac{\varepsilon(0/-)-\textit{E}_\textrm{C}}{k_B\textit{T}}] \\
\textit{e}^0_p&=\frac{1}{\textit{g}}\textit{C}_p^-\textit{N}_\textrm{V}\textrm{exp}[\frac{\textit{E}_\textrm{V}-\varepsilon(0/-)}{k_B\textit{T}}] \\
\textit{e}^+_p&=\textit{g}\textit{C}_p^0\textit{N}_\textrm{V}\textrm{exp}[\frac{\textit{E}_\textrm{V}-\varepsilon(+/0)}{k_B\textit{T}}]
\end{aligned}
\end{equation}
where $N_\textrm{C}$ and $N_\textrm{V}$ are effective density of states in the \ac{CB} and \ac{VB}, respectively. 
\textit{g} is the degeneracy factor, discussed in Refs. \cite{kavanagh_impact_2022,mosquera-lois_imperfections_2023}.

Under steady-state conditions, the net recombination is zero. 
By further considering the relation $\textit{F}^+$+$\textit{F}^0$+$\textit{F}^-$=1, the occupation functions are written as \cite{abou2011advanced}
\begin{equation}
\begin{aligned}
\textit{F}^+&=\frac{\textit{P}^0\textit{P}^-}{\textit{N}^+\textit{P}^-+\textit{P}^0\textit{P}^-+\textit{N}^+\textit{N}^0}\\
\textit{F}^0&=\frac{\textit{N}^+\textit{P}^-}{\textit{N}^+\textit{P}^-+\textit{P}^0\textit{P}^-+\textit{N}^+\textit{N}^0}\\
\textit{F}^-&=\frac{\textit{N}^+\textit{N}^0}{\textit{N}^+\textit{P}^-+\textit{P}^0\textit{P}^-+\textit{N}^+\textit{N}^0}
\end{aligned}
\end{equation}
where the variables $\textit{N}^+$, $\textit{N}^0$, $\textit{P}^-$ and $\textit{P}^0$ are defined as
\begin{equation}
\begin{aligned}
\textit{N}^+ &= \textit{n}\textit{C}_n^+ + \textit{e}_p^+ \\
\textit{N}^0 &= \textit{n}\textit{C}_n^0 + \textit{e}_p^0 \\
\textit{P}^0 &= \textit{p}\textit{C}_p^0 + \textit{e}_n^0 \\
\textit{P}^- &= \textit{p}\textit{C}_p^- + \textit{e}_n^-
\end{aligned}
\end{equation}

The net recombination rate \textit{R} for an amphoteric defect is thus written as \cite{abou2011advanced}
\begin{equation}
\begin{aligned}
\textit{R} &= r_1 - r_2 + r_5 - r_6 \\
&=\textit{N}_T(\textit{n}\textit{p}-\textit{n}_\textrm{i}^2)
\frac{\textit{C}_n^+\textit{C}_p^0\textit{P}^- + \textit{C}_n^0\textit{C}_p^-\textit{N}^+}
{\textit{N}^+\textit{P}^-+\textit{P}^0\textit{P}^-+\textit{N}^+\textit{N}^0}
\end{aligned}
\end{equation}
where \textit{N}$_\textrm{T}$ is the total concentration of the defect with all possible charge states. \textit{n} and \textit{p} are concentrations of electrons and holes, respectively. \textit{n}$_\textrm{i}$ is the intrinsic carrier concentration. $\textit{C}_{n/p}$ is the capture coefficient for electrons/holes and the superscript of capture coefficients refers to the starting charge state of the process.
The total recombination rate is the sum of recombination rates for all defect species in a material.

\begin{acknowledgement}
We acknowledge stimulating discussions on defect-mediated recombination with Sunghyun Kim.
Via our membership of the UK's HEC Materials Chemistry Consortium, which is funded by EPSRC (EP/X035859/1), this work used the ARCHER2 UK National Supercomputing Service (http://www.archer2.ac.uk).
This work was supported by the Leverhulme Trust (project RPG-2021-191).
S.R.K. acknowledges the Harvard University Center for the Environment (HUCE) for funding a fellowship.
\end{acknowledgement}

\begin{suppinfo}
Extended methods; defect thermodynamics; non-radiative carrier capture; simulated current-voltage curves.
\end{suppinfo}

\bibliography{references}

\end{document}


\begin{acronym}

\acro{PCE}{power conversion efficiency}
\acro{VASP}{Vienna Ab initio Simulation Package}
\acro{DFT}{density functional theory}
\acro{PAW}{projector augmented-wave}
\acro{SRH}{Shockley-Read-Hall}
\acro{CB}{conduction band}
\acro{VB}{valence band}
\acro{CBM}{conduction band minimum}
\acro{VBM}{valence band maximum}
\end{acronym}

\section*{S1. Computational methods}

\subsection*{Trap-limited conversion efficiency}
By accounting for both radiative and non-radiative recombination processes, the net current density \textit{J} under an applied bias voltage \textit{V} can be expressed as 
\begin{equation}
\textit{J}(V;W) = \textit{J}_\textrm{SC}(\textit{W}) + \textit{J}_0^\textrm{rad}(\textit{W})[1-\textrm{exp}(\frac{eV}{k_\textrm{B}\textit{T}})]-e\textit{R}(\textit{V})\textit{W}
\end{equation}
where \textit{W} is the film thickness. \textit{e} is the elementary charge. $\textit{J}_0^\textrm{rad}$ is the saturation current. 
The short-circuit current $\textit{J}_\textrm{SC}$ is given by \cite{kim2020upper}
\begin{equation}
\textit{J}_\textrm{SC}(W) = e\int_{E_g}^{\infty}\textit{a}(\textit{E};\textit{W})\Phi_\textrm{sun}(\textit{E})\textrm{d}\textit{E}
\end{equation}
where \textit{a} is the photon absorptivity. $\Phi_\textrm{sun}(\textit{E})$ is incident spectral photon flux density at the photon energy \textit{E}. 
~\\
~\\
The maximum efficiency is defined as the ratio of the maximum power density to the incident light power density, which is given by
\begin{equation}
\eta_{max} = \textrm{max}_V(\frac{JV}{e\int_{0}^{\infty}E\Phi_\textrm{sun}(E)\textrm{d}E})
\end{equation}

\subsection*{First-principles calculations}
All electronic structure calculations were performed using Kohn-Sham \ac{DFT} \cite{kohn1965self,dreizler1990density} as implemented in \ac{VASP} \cite{kresse1996efficient}. 
The \ac{PAW} method \cite{kresse1999ultrasoft} was employed with a converged plane-wave energy cutoff of \SI{350}{\electronvolt}.
Heyd-Scuseria-Ernzerhof hybrid exchange-correlation functional (HSE06)\cite{heyd2003hybrid} and the D3 dispersion correction\cite{grimme2004accurate} were used for both geometry optimization and total energy calculations for each defect, which have been shown to accurately reproduce the structural and electronic properties of \ce{Sb2S3}\cite{wang2022lone}.
Spin-orbit coupling (SOC) effects have been reported to have a negligible impact on \ce{Sb2S3} \cite{filip2013g} and were thus not considered in this work.
~\\
\textbf{Defect modeling.}
All point defects were simulated using a 3$\times$1$\times$1 supercell (with dimensions \SI{11.39}{\angstrom}$\times$\SI{11.21}{\angstrom}$\times$\SI{11.39}{\angstrom}) containing 60 atoms.
This supercell size has been demonstrated to be adequate for this system  \cite{zhao2021intrinsic,chen2022codoping,zhang2022sulfur}, given the relatively small charge corrections resulting from large dielectric constants of \ce{Sb2S3}\cite{wang2022lone}.
The convergence criterion of forces on each atom was set to \SI{0.01}{eV/\angstrom}.
For both geometry optimisation and static calculations, spin polarisation was turned on and a 2$\times$2$\times$2 $\varGamma$-centred \textit{k}-point mesh was used.
~\\
The ground-state defect configurations were obtained using the \textsc{doped} Python package (v0.0.7) \cite{kavanagh2024} and \textsc{ShakeNBreak}\cite{mosquera2022shakenbreak,mosquera2023identifying}.
The workflow of generating and optimizing defects follows the same procedures as described in our previous work \cite{wang2024upper}.
The initial configurations for interstitial defects were generated using the Voronoi scheme \cite{rycroft2009voro}. This method identifies corners, edges and face centers of the Voronoi polyhedra as potential distinct sites, giving rise to twelve inequivalent interstitial sites in \ce{Sb2S3}.

\section*{S2. Defect thermodynamics in \ce{Sb2S3}}
\begin{figure}[h!]
    \centering    {\includegraphics[width=0.75\textwidth]{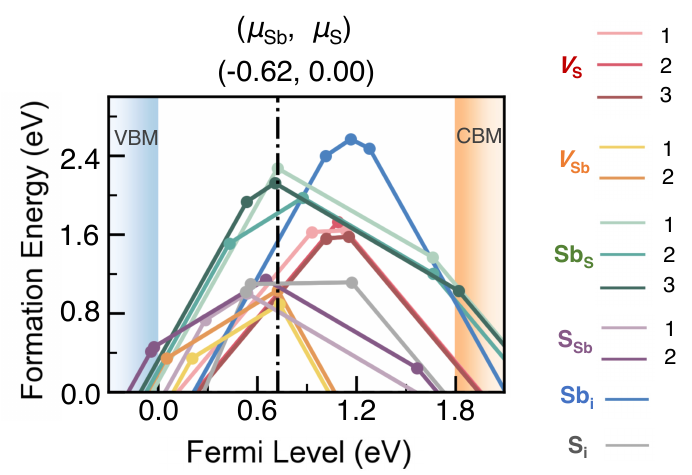}} \\
    \caption{Formation energies of all intrinsic point defects in \ce{Sb2S3} under S-rich conditions. The dashed line indicates self-consistent Fermi level at \SI{300}{\kelvin} in \ce{Sb2S3} crystals grown at \SI{603}{\kelvin} \cite{zhu2023parallel,chen2019preferentially}.}
    \label{fig_carrier_conc}
\end{figure}
\clearpage

\subsection*{S2.2 Transition energy levels of defects}
\begin{figure}[h!]
    \centering    {\includegraphics[width=\textwidth]{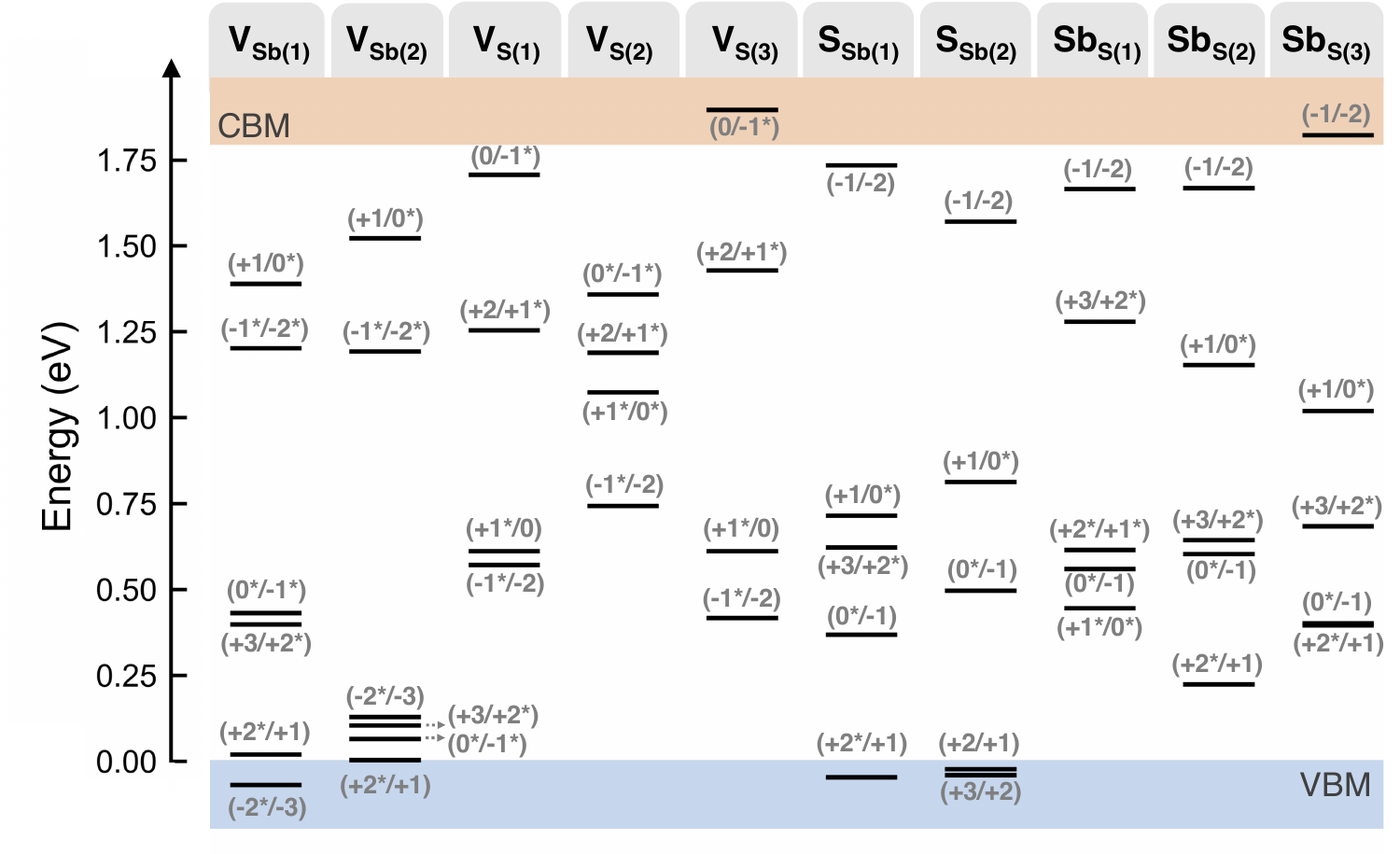}} \\
    \caption{Calculated charge state transition levels of intrinsic point defects with high concentrations in \ce{Sb2S3}. Metastable charge states are indicated with asterisks (*), and the Fermi level is referenced to the valence band maximum (VBM). }
    \label{fig_tl}
\end{figure}
\clearpage

\section*{S3. Non-radiative carrier capture processes}

\begin{figure}[H]
    \centering    {\includegraphics[width=\textwidth]{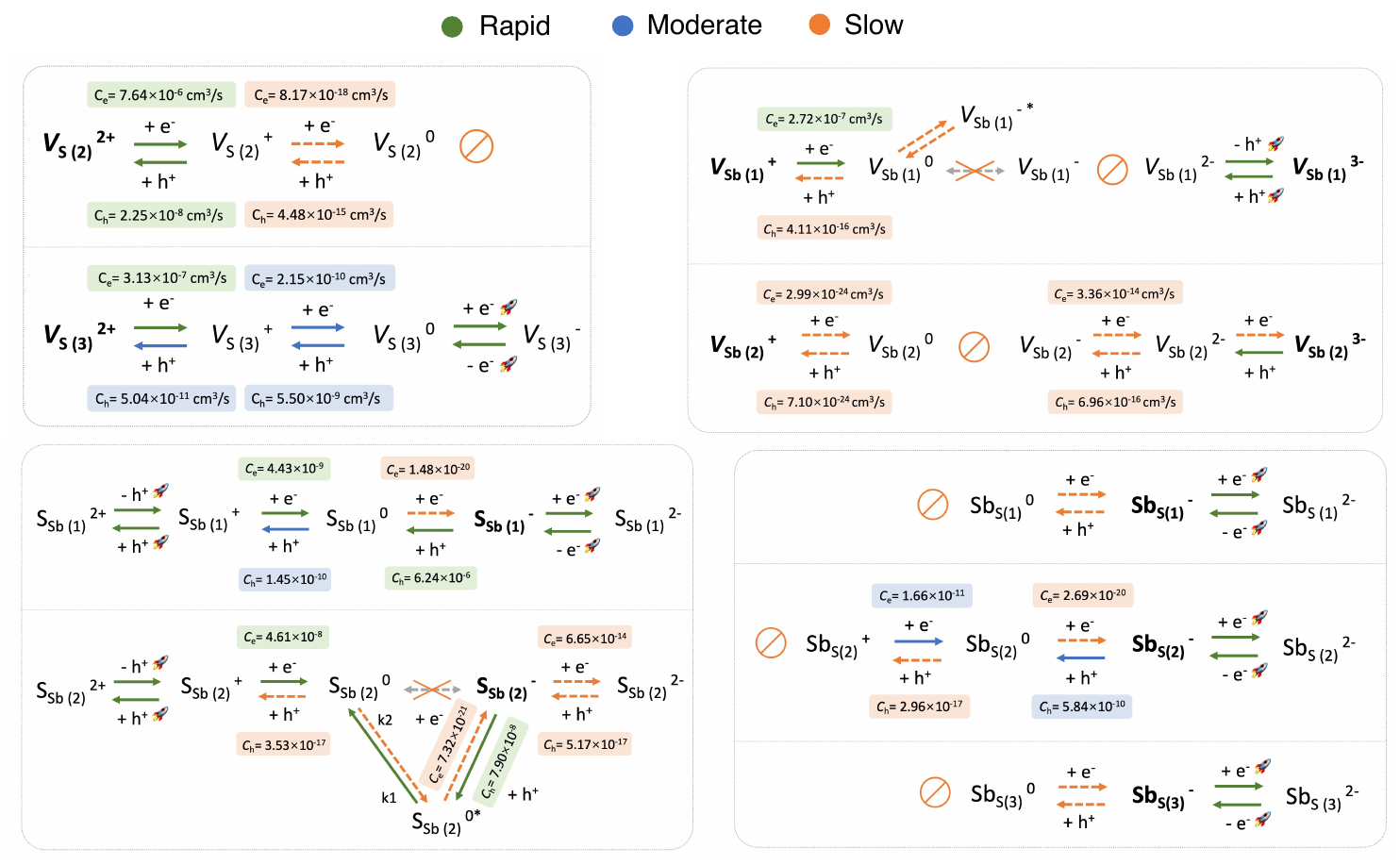}} \\
    \captionsetup{font=small, skip=10pt}
    \caption{Pathways for electron and hole capture via traps with high carrier concentrations in \ce{Sb2S3}. Defect species in bold are thermodynamically stable states at calculated self-consistent Fermi levels, which are the most likely starting points in capture processes. The defect species with superscript asterisks refer to metastable defect configurations. $C_e$ and $C_h$ are electron and hole capture coefficients, respectively. Green, blue and orange colours indicate rapid, intermediate and slow capture. `$\oslash$' refers to transitions from states with extremely low predicted concentrations under illumination. Transitions with large mass-weighted displacements are also ruled out, as indicated by an `X' mark.
    Capture coefficients smaller than \SI{e-25}{{\cubic\cm\per\second}} are not shown. 
    }
    \label{fig_path}
\end{figure}

\section*{S4. Trap-limited conversion efficiency}

\begin{figure}[H]
    \centering    {\includegraphics[width=0.8\textwidth]{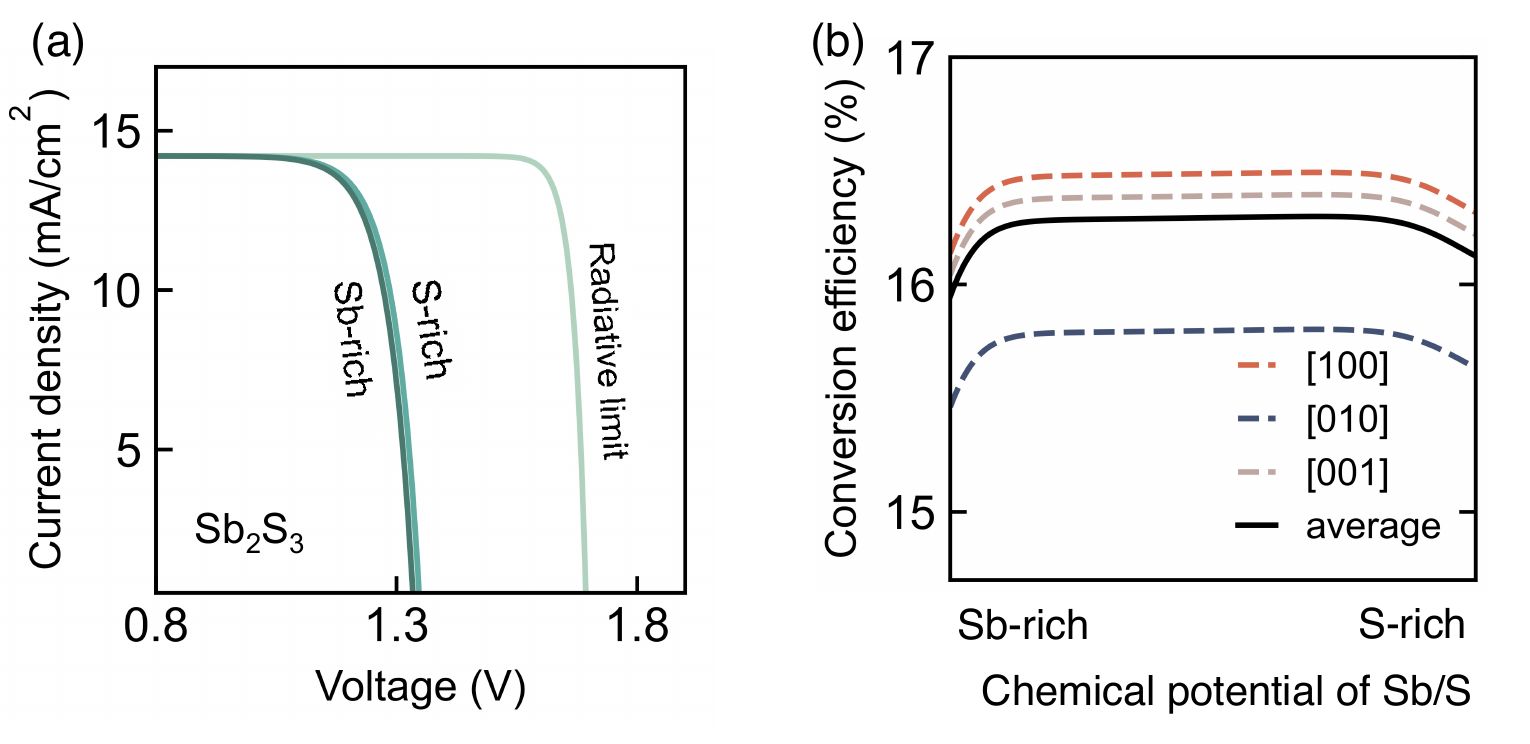}} \\
    \captionsetup{font=small, skip=10pt}
    \caption{(a) Calculated current density-voltage ($J-V$) curves for \ce{Sb2S3}, assuming the radiative limit (only band-to-band radiative recombination losses) and including defect-induced non-radiative recombination under S/Sb-rich growth conditions. The radiative limit is calculated by averaging the optical absorption coefficients along [100], [010] and [001] crystallographic directions. (b) Trap-limited conversion efficiency as a function of the growth condition. [100], [010] and [001] correspond to the crystallographic directions in \ce{Sb2S3}. All results shown correspond to a film thickness of 400 nm, and room temperature defect concentrations assuming an annealing temperature of 603 K\cite{zhu2023parallel,chen2019preferentially}.
    }
    \label{figs_pce}
\end{figure}

Fig. \ref{figs_pce}(a) shows the predicted current density-voltage (J-V) curves for \ce{Sb2S3} solar cells, incorporating both the radiative limit and trap-assisted non-radiative recombination processes. 
The radiative limit is predicted based on the calculated band gap, directionally-averaged optical absorption coefficients, and an assumed film thickness of \SI{400}{\nm} \cite{zhu2023parallel}.
The predicted short-circuit current density (\textit{J}$\mathrm{_{SC}}$) is 14.2 mA/cm$^2$ (Fig. \ref{figs_pce}(a)), which is lower than the \textit{J}$\mathrm{_{SC}}$ of 19.3 mA/cm$^2$ observed in the highest-efficiency \ce{Sb2S3} solar cell\cite{zhu2023parallel}. 
This discrepancy is attributed to the exclusion of temperature effects in the model, which would otherwise increase the calculated \textit{J}$\mathrm{_{SC}}$.
Additionally, the predicted open-circuit voltage \textit{V}$\mathrm{_{OC}}$ deficit due to radiative recombination is \SI{0.10}{/V}.
Non-radiative recombination significantly contributes to \textit{V}$\mathrm{_{OC}}$ deficit, with total predicted deficits of \SI{0.45}{/V} and \SI{0.44}{/V} under Sb-rich and S-rich conditions, respectively (Fig. \ref{figs_pce}(a)).

The upper limit to conversion efficiencies in \ce{Sb2X3} solar cells have been predicted using the TLC model \cite{kim2020upper,kim2021ab} as shown in Fig. \ref{figs_pce}(b).
The anisotropic conversion efficiency was calculated based on the corresponding optical absorption coefficients.
Our predictions indicate that the highest trap-limited conversion efficiency of \SI{16.5}{\percent} can be achieved along the [100] direction, which corresponds to the direction parallel to the quasi-one-dimensional [Sb$_4$S$_6$]$_n$ ribbons under optimal growth conditions in \ce{Sb2S3}.
However, under the same conditions, the maximum difference in efficiency along different directions is only \SI{0.69}{\percent}.

\section*{S5. Electron-phonon matrix element calculations}
\begin{figure}[h]
    \centering    {\includegraphics[width=\textwidth]{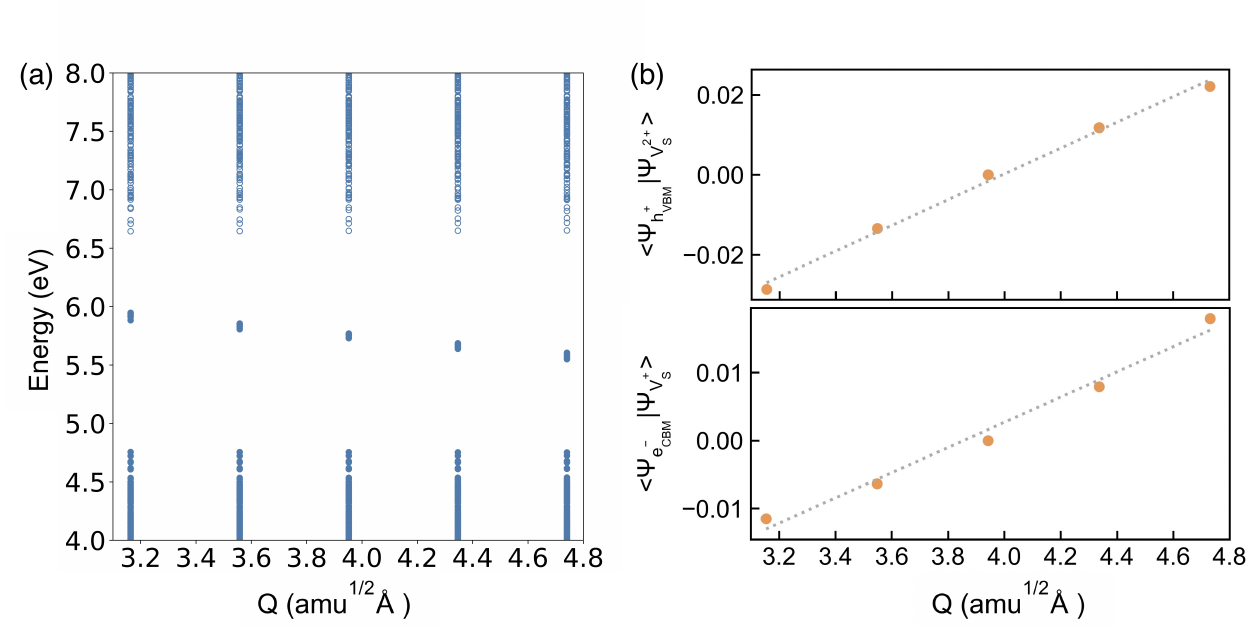}} \\
    \caption{Calculation of the electron-phonon matrix element for \textit{V}$^{2+}_\mathrm{S(2)}$ $\leftrightarrow$  \textit{V}$^{+}_\mathrm{S(2)}$ transition. (a) Eigenvalues (spin-up channel) of defect states and band edge wavefunctions as a function of the generalized coordinate\textit{Q}. (b) Overlap integral $\langle{\psi_{i}}|{\psi_{f}}\rangle$ as a function of \textit{Q} for hole and electron capture processes.}
    \label{fig_wif}
\end{figure}

The carrier capture coefficients are calculated using the electron-phonon coupling matrix element, which is expanded in a Taylor series around {\textit{Q}$_0$}, under the linear-coupling approximation \cite{alkauskas2014first}. 
Fig. \ref{fig_wif} presents the datapoints used to calculate the electron-phonon matrix element for the \textit{V}$^{2+}_\mathrm{S(2)}$ $\leftrightarrow$  \textit{V}$^{+}_\mathrm{S(2)}$ transition.
In Fig. \ref{fig_wif}(a), the eigenvalues of the spin-up channel defect state exhibit a linear dependence on the generalized coordinate \textit{Q}, while the valence band maximum (VBM) and conduction band minimum (CBM) remain constant with respect to \textit{Q}. 
Fig. \ref{fig_wif}(b) illustrates the overlap integral $\langle{\psi_{i}}|{\psi_{f}}\rangle$ as a function of \textit{Q} for both hole and electron capture processes, with linear fits applied to each.

\clearpage
\bibliography{References}